\documentclass[prl,twoside,twocolumn,superscriptaddress,showpacs]{revtex4}
\usepackage{graphicx}
\usepackage{amsmath}
\usepackage{amssymb}

\begin{document}

\title{Charge Order Superstructure with Integer Iron Valence in Fe$_2$OBO$_3$ }

\author{M. Angst}
 \email[email: ]{angst@ornl.gov}
\affiliation{Materials Science and Technology Division, Oak Ridge
National Laboratory, Oak Ridge, TN 37831, USA}
\author{P. Khalifah}
\affiliation{Department of Chemistry, University of Massachusetts,
Amherst, MA 01003, USA}
\author{R.~P. Hermann}
\affiliation{Institut f\"ur Festk\"orperforschung, Forschungszentrum
J\"ulich GmbH, D-52425 J\"ulich, Germany} \affiliation{Department of
Physics, B5, Universit\'e de Li\`ege, B-4000 Sart-Tilman, Belgium}
\author{H.~J. Xiang}
\author{M.-H.~Whangbo}
\affiliation{Department of Chemistry, North Carolina State
University, Raleigh, NC 27695, USA}
\author{V. Varadarajan}
\author{J. W. Brill}
\affiliation{Department of Physics and Astronomy, University of
Kentucky, Lexington, Kentucky 40506, USA}
\author{B.~C. Sales}
\author{D.~Mandrus}
\affiliation{Materials Science and Technology Division, Oak Ridge
National Laboratory, Oak Ridge, TN 37831, USA}

\date{\today}
\begin{abstract}
Solution-grown single crystals of Fe$_2$OBO$_3$ were characterized
by specific heat, M\"ossbauer spectroscopy, and x-ray diffraction. A
peak in the specific heat at $340\,{\rm K}$ indicates the onset of
charge order. Evidence for a doubling of the unit cell at low
temperature is presented. Combining structural refinement of
diffraction data and M\"ossbauer spectra, domains with diagonal
charge order are established. Bond-valence-sum analysis indicates
integer valence states of the Fe ions in the charge ordered phase,
suggesting Fe$_2$OBO$_3$ is the clearest example of ionic charge
order so far.
\end{abstract}

\pacs{61.50.Ks, 71.30.+h, 71.28.+d, 61.10.Nz}

\maketitle

Many physical phenomena in transition metal oxides, including
colossal magnetoresistance \cite{ChuangMilward} and high-temperature
superconductivity \cite{ReznikValla}, are related to charge ordering
(CO). Ideally, CO consists of charge carriers localizing on ions
with different integer valences forming an ordered pattern
\cite{WignerVerwey}. However, the application of this ``ionic CO''
concept has been controversial \cite{Coey04,Garcia04}, because
observed valence separations are usually small, and there is no
clear example of CO with integer valences.

\begin{figure}[tb]
\includegraphics[width=0.6\linewidth]{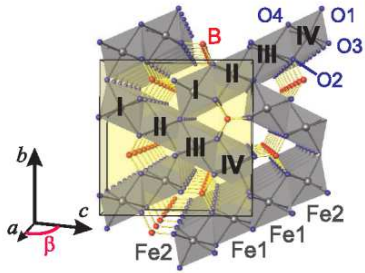}
\caption{ (Color online) Crystal structure of Fe$_2$OBO$_3$ at
$355\,{\rm K}$. The unit cell (shaded yellow) is orthorhombic
($Pmcn$), at lower $T$ the symmetry is lowered to monoclinic, with a
CO superstructure having a $2a\!\times\! b\!\times\! c$ cell.
Structurally distinct Fe$1$O$_6$ and Fe$2$O$_6$ octahedra build
ribbons of four edge-sharing chains (numbered, see text) along $a$.
Detail views of the ribbons at $355$ and $100\,{\rm K}$ are given in
Fig.\ \ref{COlowT}. } \label{struc}
\end{figure}

M\"{o}ssbauer spectra on the mixed-valent warwickite Fe$_2$OBO$_3$
suggested a large, though not quantified, Fe valence separation
below the onset of a monoclinic distortion of the structure (Fig.\
\ref{struc}) at $317\,{\rm K}$ \cite{Attfield98}. It is natural then
to suspect an ordered arrangement of Fe$^{2+}$ and Fe$^{3+}$ ions
(ionic CO), and Fe$_2$OBO$_3$ has been suggested as an example of
electrostatically driven CO \cite{Attfield98}. However, no
experimental evidence of a CO superstructure was found on the
available polycrystalline samples, and consequently the occurrence
of CO in Fe$_2$OBO$_3$ has been under debate \cite{GarciaLeonov}.

Here, we report the first observation of superstructure reflections
in single-crystalline Fe$_2$OBO$_3$, using X-ray diffraction.
Combining structural refinement, M\"ossbauer spectroscopy, and
electronic structure calculations, we establish a diagonal CO
configuration. Bond-valence-sum analysis indicates that the ordered
iron valence states are very close to integer Fe$^{2+}$ and
Fe$^{3+}$. Thus, Fe$_2$OBO$_3$ is an excellent example of ionic CO.
We discuss implications of the large structural modulations on the
relevance of the electron-lattice coupling in driving the CO.

\begin{figure}[b]
\includegraphics[width=0.95\linewidth]{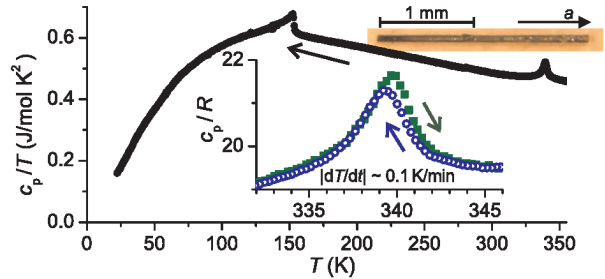}
\caption{(Color online) Specific heat $c_p / T$, measured by ac
calorimetry. Lower inset: $c_p$ around $340\,{\rm K}$. Upper inset:
Fe$_2$OBO$_3$ crystal.} \label{Figed}
\end{figure}

Needle-like single crystals of Fe$_2$OBO$_3$ (Fig.\ \ref{Figed}
inset) with length up to $1.5\,{\rm cm}$ were grown from a flux with
a procedure very similar to the growth of
Fe$_{1.91}$V$_{0.09}$OBO$_3$ reported by Balaev {\em et al.}
\cite{Balaev03}, except that we omitted V$_2$O$_3$ from the flux to
avoid V doping. $^{57}$Fe M{\"{o}}ssbauer spectra, obtained on
powdered crystals using a constant-acceleration spectrometer
\cite{Hermann04}, were similar to previous results
\cite{Attfield98,Douvalis00} with isomer shifts at low $T$ (Fig.\
\ref{BVSextra}b) indicating divalent and trivalent Fe with no
electron hopping. The specific heat (Fig.\ \ref{Figed}) was measured
with ac calorimetry \cite{Chung93} on several crystals giving
consistent results. A step-like feature around $153\,{\rm K}$ is due
to the known magnetic transition \cite{Continentino01}. A peak at
$340\,{\rm K}$ with $\sim\! 0.25\,{\rm K}$ hysteresis suggests an
additional, weakly first-order, phase transition, which according to
a powder diffraction study \cite{note_Payzant} corresponds to the
monoclinic-orthorhombic transition, and which we attribute to to the
onset of CO. The estimated entropy change associated with the
transition is very small ($\sim\!0.01\,R/$Fe ion, $R=8.14\,{\rm
J/mol/K}$), but we point out that ac calorimetry is sensitive only
to the reversible heat flow.

\begin{figure}[t]
\includegraphics[width=0.96\linewidth]{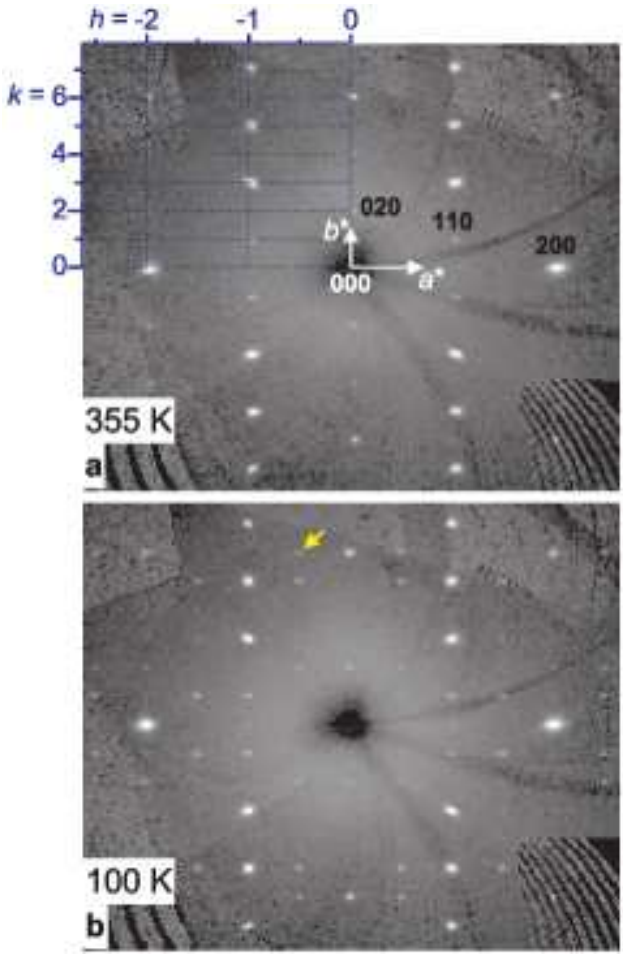}
\caption{(Color online) Composite X-ray diffraction precession
images with intensities extracted from about $500$ individual
frames. a: $355\,{\rm K}$. Due to an $n$ glide plane, spots with
$h+k$ odd are systematically absent. b: $100\,{\rm K}$. Weak
additional spots (one indicated by an arrow) index to
$(h\!+\!\frac{1}{2},k,0)$, indicating a superstructure with doubled
($2a\times b\times c$) unit cell.}
\label{Figprec}
\end{figure}

\begin{table}[!t]
\begin{center}
\caption{\small Refinement results \cite{noteICSD} for Fe sites in
Fe$_2$OBO$_3$ at $355\,{\rm K}$ (space group $Pmcn$, $a\!=\!
3.18\,{\rm \AA}$, $b\!=\! 9.40\,{\rm \AA}$, $c\!=\! 9.25\,{\rm
\AA}$) and $100\,{\rm K}$ ({\em average} structure, space group
$Pc$, $2a\!=\!6.33\,{\rm \AA}$, $b\!=\! 9.38\,{\rm \AA}$, $c\!=\!
9.25\,{\rm \AA}$). Average Fe-O bondlengths, bond-valence-sum, and
distortion parameter (see text). \label{BVStab} }
\begin{tabular}{llll}
\hline \hline
Site   & $\ \ \ \langle d$(Fe-O)$\rangle /{\rm \AA}$ &   $\ \ \ \ $Bond Valence Sum  & $\ \ \ \ $Distortion$\ \Gamma$   \\
\hline

\multicolumn{4}{c}{$355\,{\rm K}$}\\

\hline
Fe1    &   $\ \ \ \ \ \ \ $2.094(2)    &   $\ \ \ \ \ \ \ \ \ \ \ \ $2.39(1)  & $\ \ \ \ \ \ \ \ \ \ \ $2.8 \\
Fe2    &   $\ \ \ \ \ \ \ $2.088(2)    &   $\ \ \ \ \ \ \ \ \ \ \ \ $2.46(1)  & $\ \ \ \ \ \ \ \ \ \ \ $4.0 \\
\hline

\multicolumn{4}{c}{$100\,{\rm K}$}\\

\hline
Fe1$a$ &   $\ \ \ \ \ \ \ $2.140(6)    &   $\ \ \ \ \ \ \ \ \ \ \ \ $2.04(3)  & $\ \ \ \ \ \ \ \ \ \ \ $5.9 \\
Fe2$a$ &   $\ \ \ \ \ \ \ $2.154(6)    &   $\ \ \ \ \ \ \ \ \ \ \ \ $2.03(3)  & $\ \ \ \ \ \ \ \ \ \ \ $7.5 \\
Fe1$b$ &   $\ \ \ \ \ \ \ $2.020(6)    &   $\ \ \ \ \ \ \ \ \ \ \ \ $3.02(5)  & $\ \ \ \ \ \ \ \ \ \ \ $0.4 \\
Fe2$b$ &   $\ \ \ \ \ \ \ $2.019(6)    &   $\ \ \ \ \ \ \ \ \ \ \ \ $3.07(5)  & $\ \ \ \ \ \ \ \ \ \ \ $0.0 \\
Fe1$c$ &   $\ \ \ \ \ \ \ $2.098(7)    &   $\ \ \ \ \ \ \ \ \ \ \ \ $2.35(8)  & $\ \ \ \ \ \ \ \ \ \ \ $2.9 \\
Fe2$c$ &   $\ \ \ \ \ \ \ $2.087(6)    &   $\ \ \ \ \ \ \ \ \ \ \ \ $2.47(8)  & $\ \ \ \ \ \ \ \ \ \ \ $3.2 \\
Fe1$d$ &   $\ \ \ \ \ \ \ $2.098(6)    &   $\ \ \ \ \ \ \ \ \ \ \ \ $2.34(8)  & $\ \ \ \ \ \ \ \ \ \ \ $2.9 \\
Fe2$d$ &   $\ \ \ \ \ \ \ $2.086(6)    &   $\ \ \ \ \ \ \ \ \ \ \ \ $2.48(8)  & $\ \ \ \ \ \ \ \ \ \ \ $3.6 \\
\hline \hline
\end{tabular}
\end{center}
\end{table}

\begin{figure*}[thb]
\includegraphics[width=0.93\linewidth]{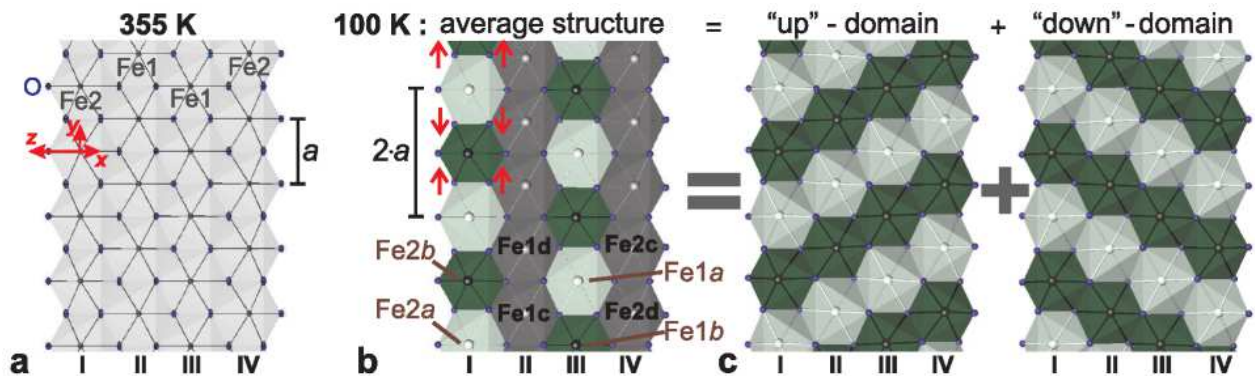}
\caption{ (Color online) Charge order in the four-chain ribbons
(numbered as in Fig.\ \ref{struc}). a: $355\,{\rm K}$. Atoms are
drawn as thermal ellipsoids. For one FeO$_6$ octahedron a local
coordinate system is indicated. b,c: CO at $100\,{\rm K}$ (high/low
valence Fe is shaded dark/bright): The ``global'' structure, refined
in $Pc$ (b) arises from averaging domains with different diagonal
order (c). Large oxygen shifts within the superstructure are
emphasized by red arrows in b. } \label{COlowT}
\end{figure*}

\begin{figure}[b]
\includegraphics[width=0.73\linewidth]{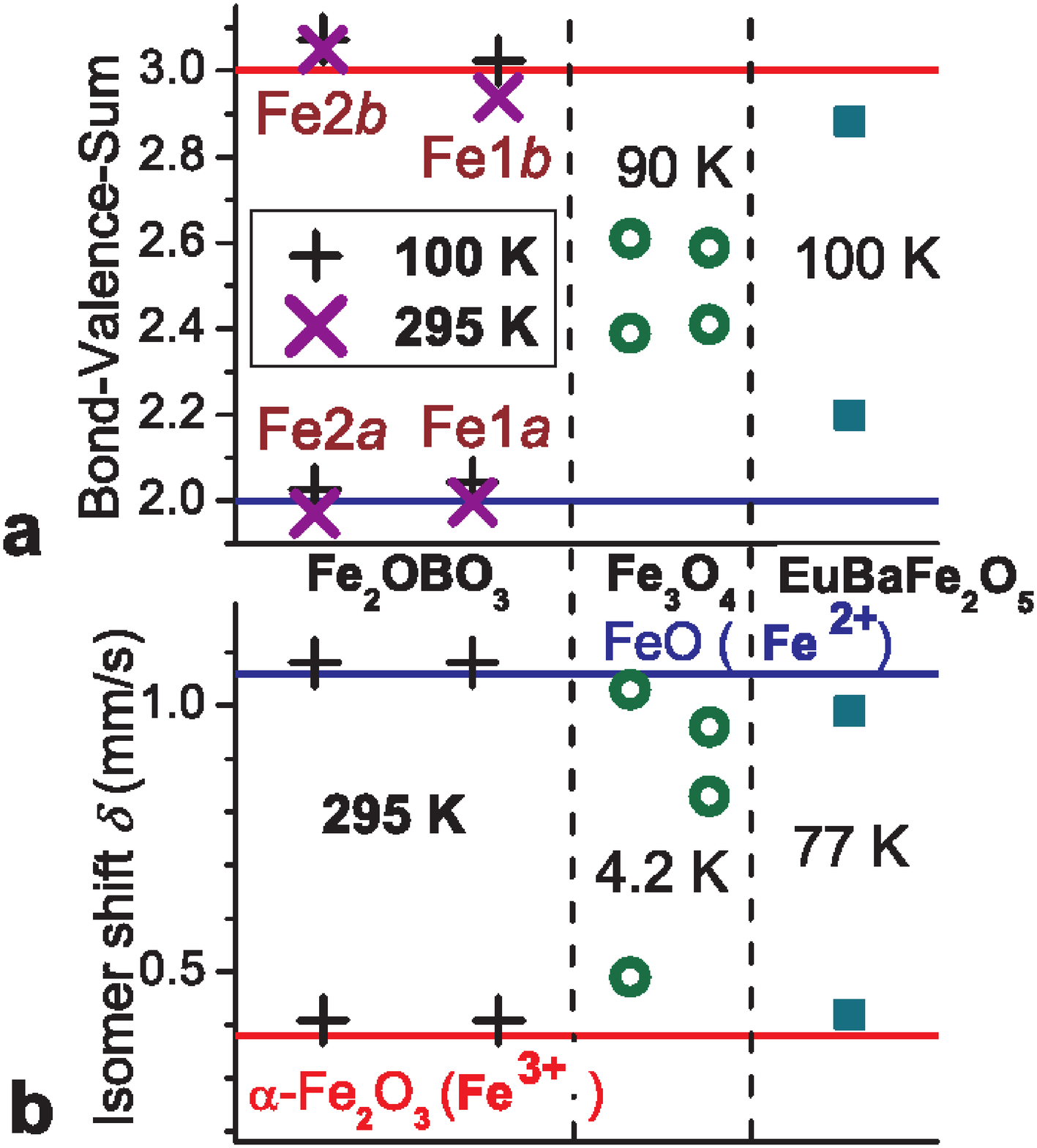}
\caption{ (Color online) Integer valence separation. a: Fe valence
from bond-valence-sum analysis on four sites indicated in Fig.\
\ref{COlowT}b ($+$), compared to the classical CO example magnetite
(B sites, $\circ$\cite{Wright01}) and the clearer example
EuBaFe$_2$O$_5$ ($\blacksquare$\cite{Karen07}). b: M{\"{o}}ssbauer
isomer shifts relative to $\alpha$Fe ($+$), compared to magnetite (B
sites, $\circ$\cite{Berry98}) and EuBaFe$_2$O$_5$
($\blacksquare$\cite{Karen07}). The two horizontal lines represent
isomer shifts\cite{Shirane62} for Fe$^{2+}$ in FeO (blue) and
Fe$^{3+}$ in Fe$_2$O$_3$ (red) at room temperature. }
\label{BVSextra}
\end{figure}

To corroborate the CO, we collected single-crystal X-ray diffraction
data with a Bruker SMART APEX CCD diffractometer, using Mo$K\alpha$
radiation. Figure \ref{Figprec} shows composite precession images of
the $hk0$ plane with intensities extracted from about $500$
individual frames measured at $355\,{\rm K}$ (a) and $100\,{\rm K}$
(b). The pattern at $355\,{\rm K}$ is expected for the $Pmcn$ (no.\
62) space group, and the refined structure (Figs.\ \ref{struc} and
\ref{COlowT}a \cite{noteICSD}) was very similar to the one reported
for polycrystalline Fe$_2$OBO$_3$ \cite{Attfield98}. At $100\,{\rm
K}$, weak additional spots indexing to $(h\!+\!\frac{1}{2},k,0)$
indicate a superstructure with $2a\!\times\! b\!\times\! c$ cell,
attributed to CO as detailed below. Visual inspection of the raw
data indicated overlapping peaks of roughly equal intensity
consistent with monoclinic twinning with similar weight of domains
with opposite sense of monoclinic distortion. The twinned peaks were
not sufficiently resolved and therefore the refinement was conducted
for the combined peaks. As a result, the monoclinic angle $\beta$
could not be refined, and was set to $90^{\circ}$.

To assess possible symmetries, the data were first refined with no
symmetry constraints (space group $P1$). A $Pc$ (No.\ $7$) space
group symmetry was apparent, and a corresponding refinement (Fig.\
\ref{COlowT}b) had a residual $R[F^2\!\!>\!\!4\sigma
(F^2)]\!=\!4.62\%$ significantly better than alternative space
groups \cite{noteICSD}. Small (very weak intensity) reflection
condition violations and a Flack parameter close to $50$\% suggest
that the structure is not homogeneous, but an average of domains
with different, possibly centro-symmetric, structure. Room
temperature diffraction data are very similar to the $100\,{\rm K}$
data, indicating negligible influence of the magnetic transition on
the CO.

For octahedrally coordinated Fe$^{2+}$ and Fe$^{3+}$ ions, the
expected average Fe-O bondlengths are $2.16$ and $2.02\,{\rm \AA}$,
respectively \cite{Shannon76}. For the refined structure, Fe-O
bondlengths for half of the eight Fe sites are intermediate and
similar as at $355\,{\rm K}$ (Table \ref{BVStab}), suggesting CO in
only half of the chains. However, M\"ossbauer spectra show that the
{\em local} structure is different because there is no intermediate
valence Fe in the low $T$ phase (Fig.\ \ref{BVSextra}b). The refined
structure thus arises from averaging local structures of two types
of domains. In two of the chains, II and IV in Fig.\ \ref{COlowT},
structural distortions are averaged out, indicating that these sites
have opposite valence, and thus distortions, in the two domain
types. The CO in the other two chains is clearly preserved,
indicating that their Fe sites have identical valences in all
domains. Because the Fe sites with equal $a$ position in the chains
with globally preserved CO have opposite valence, the local CO
configuration is diagonal, with ``up'' and ``down'' diagonals making
up the two types of domains (Fig.\ \ref{COlowT}c).

To corroborate the local diagonal CO, we performed first principles
electronic structure calculations (GGA+U, with fixed cell, but in
contrast to \cite{Leonov05} optimized atom positions), using
techniques as described in \cite{Xiang07}. The GGA+U calculations
($U\!=\!5.5\,{\rm eV}$, $J\!=\!0.89\,{\rm eV}$ \cite{Leonov05})
started with the crystal structure as reported by Attfield {\em et
al.}\ \cite{Attfield98} with doubled $a$ axis, in which all Fe atoms
have a uniform valence Fe$^{2.5+}$. As in \cite{Leonov05}, a
diagonal configuration with charge separation was obtained even with
fixed atom positions. Relaxing atom positions led to an additional
decrease in energy (total CO gain $174.5\,{\rm meV}/{\rm Fe}$ ion),
with CO distortions of the optimized atom positions (shown in Fig.\
\ref{COlowT}c) qualitatively equal to the refined ones (neglecting
the averaged out sites in chains II and IV), though about 20\% less
in magnitude.

The shifts in the oxygen positions (Fig.\ \ref{COlowT}b arrows)
cause the average Fe-O bondlengths for sites Fe1$a$ and Fe2$a$ to
increase and for sites Fe1$b$ and Fe2$b$ to decrease. Valence states
and ion sizes are intimately related, allowing the length of the
Fe-O bonds to be used to calculate the Fe valence through
bond-valence-sum (BVS) analysis \cite{BrownBrese}. The BVS is
\begin{equation}
V = \sum_i \exp  \left[ \left( d_0 - d_i \right) / 0.37 \right] ,
 \label{BVSformula}
\end{equation}
where $V$ is the valence of an ion to be determined, $d_i$ are the
bondlengths to other ions, and $d_0$ is a tabulated
\cite{BrownBrese} empirical parameter characteristic for a
cation-anion pair. At all temperatures the BVS of B and O are close
to $3$ and $-2$, respectively, as expected. A complication for Fe is
that the empirical $d_0$ are slightly different for different
valence states. Using tabulated \cite{BrownBrese} $d_0$ for both
Fe$^{2+}$-O and Fe$^{3+}$-O, the calculated valences for ions
Fe1$a$,Fe2$a$ and Fe1$b$,Fe2$b$ are within $0.2$ of $2$ and $3$,
respectively. As is customary in this case \cite{note_Rodr}, the
final valences were then calculated using $d_0$ of Fe$^{2+}$-O for
ions Fe1$a$ and Fe2$a$ and $d_0$ of Fe$^{3+}$-O for ions Fe1$b$ and
Fe2$b$. For the averaged sites in chains II,IV at low $T$ and all Fe
sites at high $T$ the BVS are intermediate between $2$ and $3$, and
the final valences were obtained by averaging the valences
calculated with $d_0$ for Fe$^{2+}$ and Fe$^{3+}$.

Averaging over anti-phase domains can only decrease the difference
between high and low valence Fe-O bondlengths in the average
structure. Consequently the valence separation from BVS obtained
from any refined global average structure provides a lower limit for
the actual local separation. We focus on the sites in chains I and
III, which provide the most stringent limit. These Fe valences are
integer within the resolution of the method (Fig.\ \ref{BVSextra}a).
Fe$_2$OBO$_3$ is the first CO oxide for which BVS clearly indicate
integer valence for both valence states. The valence separation is
considerably larger than in the so far clearest examples,
YBaFe$_2$O$_5$ \cite{Woodward03} and related compounds ($\sim\!
0.7$), and much larger than in the classical, though not fully
understood, CO example magnetite ($\lesssim \! 0.4$ \cite{Wright01})
or in the colossal magnetoresistance manganites ($\sim \! 0.45$
\cite{Garcia04}). {\em The large valence separation suggests that
Fe$_2$OBO$_3$ is an ideal example of ionic charge order}.

In agreement with the large valence difference, M{\"{o}}ssbauer
spectral isomer shifts $\delta$ (Fig.\ \ref{BVSextra}b) for two
doublets are close to $\delta$ of the Fe$^{3+}$ compound Fe$_2$O$_3$
and for the other two doublets to $\delta$ of the Fe$^{2+}$ compound
FeO. The reason that Fe$_2$OBO$_3$ is unique among oxides in
demonstrating ionic CO may be attributed to an antagonistic
inductive effect\cite{Menil85} from the B-O bonds. Since B is more
electronegative than Fe, O prefers to share its electrons with B,
making the Fe-O bonds more ionic.

The large ($>0.2\,{\rm \AA}$) oxygen shifts along $a$ accompanying
the CO also affect the distortions of the coordination octahedra. At
high $T$ the largest distortion 
is an elongation along $a$ (Fig.\ \ref{COlowT}a), leading to a
distortion parameter $\Gamma$, defined as the difference in \%
between bondlengths in the local $xy$ plane (Fig.\ \ref{COlowT}a)
and those perpendicular to it, of $3$ to $4\%$. In the CO state,
this distortion is reduced almost completely for Fe$^{3+}$ sites,
but increased for Fe$^{2+}$ sites, consistent with the extra
electron occupying a $d_{xy}$ orbital with lobes $\|a$. Because the
distortion lifts the degeneracy of the $t_{2g}$ orbitals, an energy
gain associated with the Jahn-Teller effect \cite{Jahn37} likely
contributes to the overall energy gain of charge localization and
order. Thus, CO in Fe$_2$OBO$_3$ is not driven {\em entirely} by
electrostatic interactions between carriers, although the latter
alone seems to be sufficient in establishing CO, as indicated by the
GGA+U calculations with atom positions fixed. This exemplifies the
relevance of electron-lattice effects in correlated oxides, which
likely is ubiquitous.

In summary, superstructure reflections at in Fe$_2$OBO$_3$ arise
from diagonal CO with two domains and Fe valence states very close
to integer, suggesting that Fe$_2$OBO$_3$ is an ideal example for
ionic CO. Magnitude and form of the atomic displacements in the CO
suggest that electrostatic energy is not the only relevant energy
scale, but coupling to the lattice is important as well. Very
anisotropic displacement parameters of the oxygen atoms already at
$355\,{\rm K}$ (Fig.\ \ref{COlowT}a) hint at significant precursor
effects to the CO transition; a detailed study of the evolution of
the CO with temperature is in progress.

We thank A. Payzant, W. Schweika, A. W. Sleight, B. Chakoumakos, J.
Tao, F. Grandjean, O. Swader, and O. Garlea for assistance and
discussions. Research at ORNL sponsored by the Division of Materials
Sciences and Engineering, Office of Basic Energy Sciences (OS), US
Department of Energy (DOE) (contract DE-AC05-00OR22725); at NCSU by
OS, DOE (DE-FG02-86ER45259); at UK by NSF (DMR-0400938); at UL: FNRS
credit 1.5.064.05.

\newcommand{\noopsort}[1]{} \newcommand{\printfirst}[2]{#1}
  \newcommand{\singleletter}[1]{#1} \newcommand{\switchargs}[2]{#2#1}


\begin{thebibliography}{10}
\bibitem{noteICSD}
Details of the refined structures are deposited in the Inorganic
Crystal
  Structure Database with accession numbers $417971,417972$ ($355,100\,{\rm
  K}$).
\end{thebibliography}
\end{document}